# Cavity-Enabled Self-Electro-Optic Bistability in Silicon Photonics

Arka Majumdar[1] and Armand Rundquist[2]

[1] Electrical Engineering, University of Washington, Seattle, WA-98195

[2] E. L. Ginzton Laboratory, Stanford University, Stanford, CA-94305

**ABSTRACT:** We propose a new type of bistable device for silicon photonics, using the self-electro-optic effect within an optical cavity. Since the bistability does not depend on the intrinsic optical nonlinearity of the material, but is instead engineered by means of an optoelectronic feedback, it appears at low optical powers. This bistable device satisfies all the basic criteria required in an optical switch to build a scalable digital optical computing system.

**TEXT:**

**Introduction:** Significant progress in silicon photonics in the last decade has helped to bring the long-standing goal of optical interconnects closer to a reality. With several large companies working on developing optical interconnects for data-centers, we can surmise that in the near future, such optical interconnects will proliferate throughout the computing sphere. At that point, the excess latency and energy consumption incurred in electro-optic signal conversions will be increasingly prohibitive, and so it will become necessary to perform computations directly on the optical signal. Hence, in the very near future, even though optical computing will not be a competitive technology to the existing digital computers for all applications, an increase in optical interconnects will necessitate such optical computing for specific tasks.

In general, a digital computer requires two stable states to code one bit of information. This is achieved in an optical system by exploiting optical bistability[1]. However, a simple demonstration of optical bistability is not sufficient for building an optical computing system. In fact, several approaches to optical computing have been reported in the literature, but few of them satisfy all the basic criteria that are essential to build a scalable and useful digital computing system[2]. One of those few successful approaches is an optical switch based on the self-electro-optic bistable effect[3]. In a self-electro-optic device, one engineers a positive opto-electronic feedback by using a photo-detector and an electro-optic modulator, which together generate the optical bistability. We emphasize that the development of optical interconnects will require the perfection of both electro-optic modulators and photo-detectors, which are precisely the components necessary to build a high-performance self-electro-optic device. Thus, we can expect that as the technology for building optical interconnects improves, such gains will also be realized in the quality of self-electro-optic devices. Additionally, a system based on self-electro-optic devices provides a homogeneous architecture, as the basic components of the computing device can also be used independently as modulators and detectors, greatly simplifying the task of system integration.

Previous self-electro-optic effect devices, however, were primarily based on quantum confined Stark effects in quantum wells fabricated in III-V materials. The switching energy of these devices were on the order of hundreds of picojoules, much larger than an electronic switch[4]. Apart from the large switching energy, building a system out of III-V materials also presents a challenge. In the current work, we analyze the possibility of realizing such self-electro-optic effects in silicon photonics, which is potentially a much more scalable platform. With silicon photonics going fabless[5], there is a great opportunity for innovation in designing and building a large-scale integrated optical system. A straightforward approach to realize the self-electro-optic effect would be to use a Si-Ge quantum well and the resulting quantum-confined Stark effect[6]. However, the growth of this material system is complicated. Here, we propose a new way to observe the self-electro-optic effect in a silicon photonics platform using ring cavities (embedded in a p-i-n diode) with an integrated photodetector. The photodetector can be realized in silicon photonics either by doping a specific region in the ring and using defect-mediated sub-bandgap photo-detection[7], or by depositing a layer of absorbing material on silicon, such as graphene[8-10] or III-V materials[11]. We note that, with the incorporation of a high quality factor cavity, the switching energy can also be greatly reduced.

**Generic bistability in a cavity:** Before we explain the self-electro-optic bistability in detail, let us first analyze a generic optically bistable system based on a cavity (Figure 1a). The dynamics of a cavity driven by an external laser can be described by the equation:

$$\frac{da(t)}{dt} = i\Delta a(t) - [\gamma_c + \gamma_l]a(t) + i\sqrt{2\gamma_c}S_{in}$$

where $a(t)$ denotes the intra-cavity field; $\Delta$ is the detuning of the laser from the cavity resonance; $\gamma_c$ is the coupling rate of the cavity to the outside channel (e.g., a wave-guide); $\gamma_l$ is the loss in the cavity; $S_{in}$ denotes the amplitude of the input light field. Here we assume that the cavity does not have any additional radiative loss, and the total loss in the cavity is solely due to an absorbing medium. The output field is given by $S_{out} = S_{in} + i\sqrt{2\gamma_c}a$. In the steady state, the input, output and absorbed power are, respectively, $P_{in} = |S_{in}|^2$, $P_{out} = \frac{(\gamma_c - \gamma_l)^2 + \Delta^2}{(\gamma_c + \gamma_l)^2 + \Delta^2}P_{in}$ and $P_{abs} = \frac{4\gamma_c\gamma_l}{(\gamma_c + \gamma_l)^2 + \Delta^2}P_{in}$. A feedback loop in this system can be realized by shifting the cavity resonance in response to the absorbed power as $\Delta = \Delta_o + \eta P_{abs}$. With that feedback, the steady state of the device is governed by the equation:

$$\eta^2 P_{abs}{}^3 + 2\eta\Delta_o P_{abs}{}^2 + [(\gamma_c + \gamma_l)^2 + \Delta_o{}^2]P_{abs} - 4\gamma_c\gamma_l P_{in} = 0$$

By solving this equation, we find a bistable behavior of the output power as a function of the input power (Figure 2a); that is, for a single input power, there are two possible output powers. Note that the bistable operation exists as long as $\eta\Delta_o < 0$, which causes the detuning of the cavity from the laser to change sign due to the feedback. In most optical bistabilities reported in silicon, this feedback is provided by two-photon absorption and the subsequent free carrier dispersion[12-14]. Another way to achieve the feedback is via a thermo-optic mechanism[15]. Significant progress has also been made in the bistability of a III-V photonic crystal cavity, based on electro-absorption and the free carrier effect[16, 17]. However, all of these mechanisms provide little control to the user, and in general require high optical powers. We would also like to point out that the optical power required to observe the bistability is inversely proportional to η, and by changing η one can change that power by orders of magnitude. By contrast, this optical power is not very strongly dependent on the other parameters, such as the detuning or cavity losses.

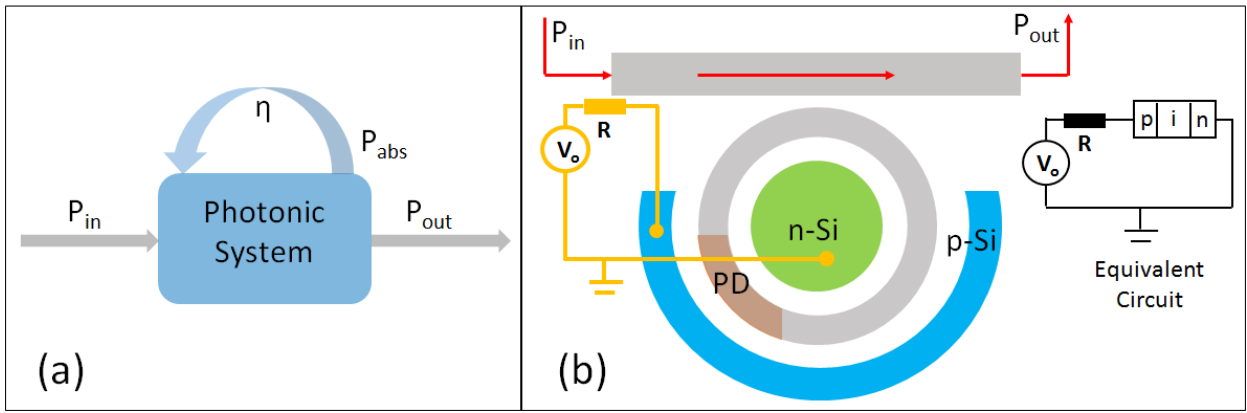

*Figure 1: (a) Block diagram for a generic optically bistable device. The bistability is caused by a positive feedback. (b) Schematic of a self-electro-optic bistable device, conceived in a silicon photonics platform: a ring cavity in a p-i-n diode with an embedded photo-detector (PD).*

**Self-electro-optic effect:** In the self-electro-optic effect, we use an explicit optoelectronic feedback method to allow the absorbed power to change the cavity resonance, which provides the user with the ability to control the strength of the feedback. This method also does not depend on the intrinsic material optical nonlinearity, which is very weak in silicon, and hence has the potential to greatly reduce the optical power required to observe bistabilty based on nonlinear optical effects. To implement the self-electro-optic effect in silicon photonics, we propose to use a ring resonator fabricated in the intrinsic region of a p-i-n diode (Figure 1b). This is the most common configuration for an electro-optic modulator based on a ring resonator[18, 19]. In this ring, we can create a photo-detector region by any of the methods mentioned earlier. We note that a ring resonator can also be described by a steady-state formalism, involving the coupling $\kappa$ between the bus waveguide and the ring, the direct transmission $t$ through the waveguide (satisfying $|t|^2 + |\kappa|^2 = 1$) and the round-trip attenuation $\mu$ of the light field inside the ring[20]. Although this formalism is well-suited for experiments, as $t$ and $\mu$ can be easily measured, it makes the dynamic behavior of the ring resonator difficult to solve. However, it can be shown that close to a single resonance, the usual steady-state description and the dynamical equation presented here converge (see the supplementary material).

For simplicity, we assume that the ring does not have any radiative loss, and that most of the loss is coming from absorption in the PD. The absorbed power in the PD gives rise to a photo-current $I_{ph} = SP_{abs}$, with $S$ being a constant. This current flows in the circuit causing the voltage across the cavity to change to $V = V_o - RSP_{abs}$. This change in the voltage can be used to change the cavity resonance. Let us first analyze a hypothetical case, where the change in the cavity detuning is linearly proportional to the applied voltage:

$$\Delta = \Delta_o + \beta(V_o - RSP_{abs}) = \Delta_o' - \beta RSP_{abs}$$

where $\beta$ is a proportionality constant which depends on the specific nanophotonic cavity parameters. This results in the same cubic equation analyzed earlier, with $\eta = -\beta RS$, which gives rise to a bistable operation (Figure 2b). Note that the bistable operation happens if $\eta < 0$, signifying a positive feedback. A similarly engineered negative, on the other hand, can be used to stabilize the ring resonators against thermal fluctuation[21]. As the bistability is accompanied by a change in the voltage across the diode, resulting in a change to the cavity resonance, we also plot the change in the voltage across the p-i-n junction (Figure 2c) and the change in the cavity detuning (Figure 2d) as a function of the input laser power. We note that the cavity resonance has to change by roughly 1 nm to observe the bistability. Although such shifts are larger than the tuning (~0.2nm) achievable with current silicon photonic modulators[19, 22], their performance can be improved by increasing their phase efficiency.

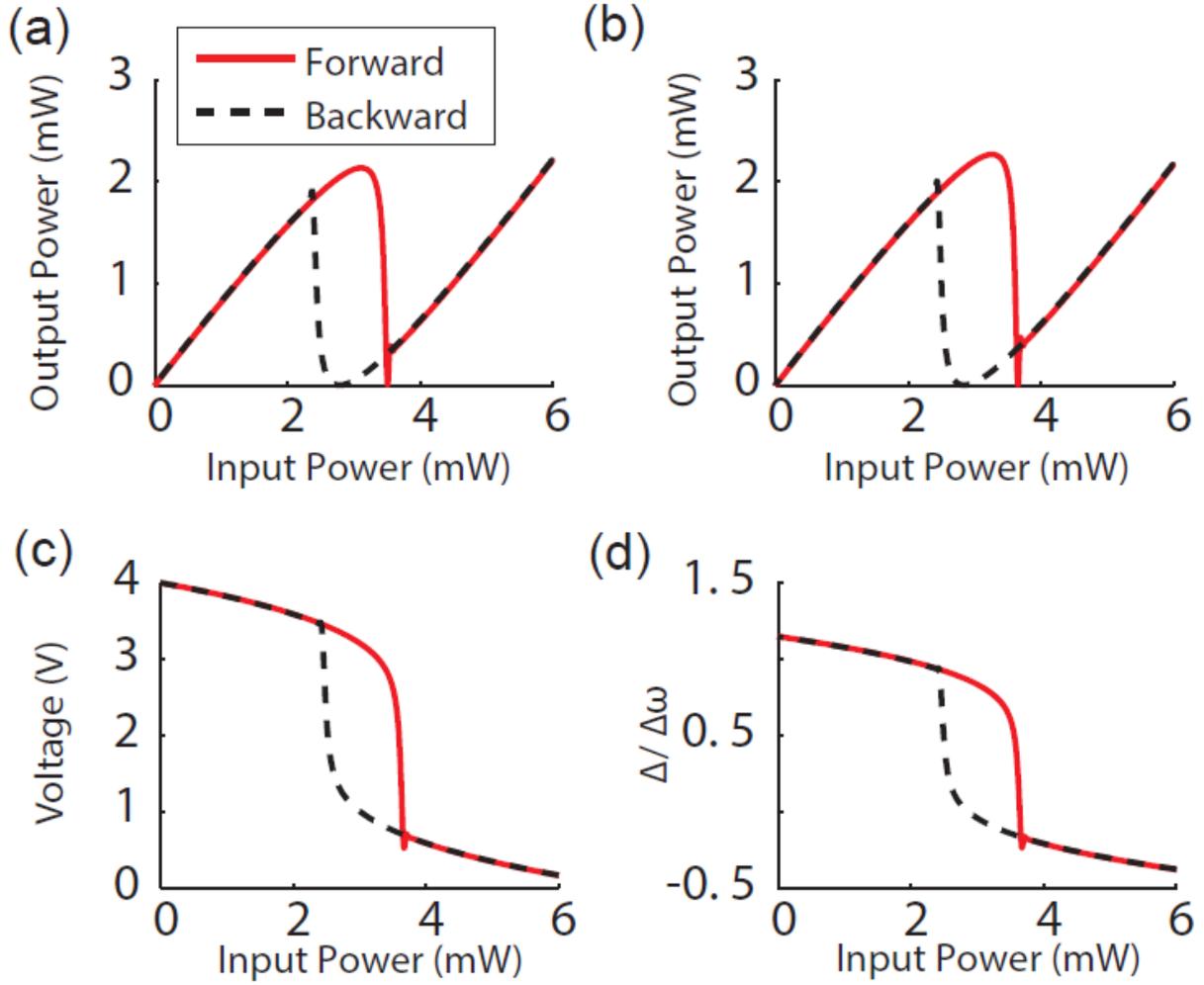

*Figure 2: (a) Input-output relation for an optically bistable device. We note that there is a hysteresis, a characteristic of the bistability. (b) Input-output relation for a self-electro-optic bistable device, assuming a linear change in the cavity resonance as a function of the voltage. The parameters for both simulations are: $\frac{\Delta_o}{2\pi} = -45$ GHz; $\frac{\gamma_c}{2\pi} = 10$ GHz and $\frac{\gamma_l}{2\pi} = 10$ GHz. For the self-electro-optic device, we assumed $V_o = 4$ V; $\alpha = 100$; $R = 1000$ Ω and $S = 1 A/W$. With these parameters the feedback constant becomes $\eta = 10^5$, which is identical to the value used in (a). The remaining plots show the change in voltage across the diode containing the ring (c) and the change in ring resonance (d), both as functions of the input laser power.*

Although we assumed a linear change in the cavity resonance frequency with the applied voltage, in a realistic device the linearity does not hold true. The applied voltage will cause a change in the width of the depletion region of the p-i-n junction as $W_D = \left(\frac{2\varepsilon_o \varepsilon_r (V+V_{bi})}{qN_A}\right)^{1/2}$. This results in an effective change of the carrier concentration of $N_A$ over a distance $W_D$. This change in the carrier concentration will change the refractive index[18] as $\Delta n = -\frac{q^2 \lambda_o^2}{8\pi^2 c^2 \varepsilon_o n m_e} \Delta N$. The cavity resonance is linearly proportional to the change in the dielectric permittivity of the material. Hence, combining everything we can write

$$\Delta = \Delta_o + \alpha \sqrt{V_o - RSP_{abs} + V_{bi}}$$

where α is again a proportionality constant that depends on the particular nanophotonic cavity parameters, more specifically the overlap of the cavity-confined electromagnetic field with the area where the carriers are changing. Figure 3 shows the simulation results under the realistic condition of carrier modulation by changes to the depletion width. We clearly observe optical bistability in this case as well. Note that in the simulation we assumed S=1A/W, which is a rather large responsivity, achievable with a graphene-based photodetector. However, the sensitivity can be increased by using a trans-impedance amplifier.

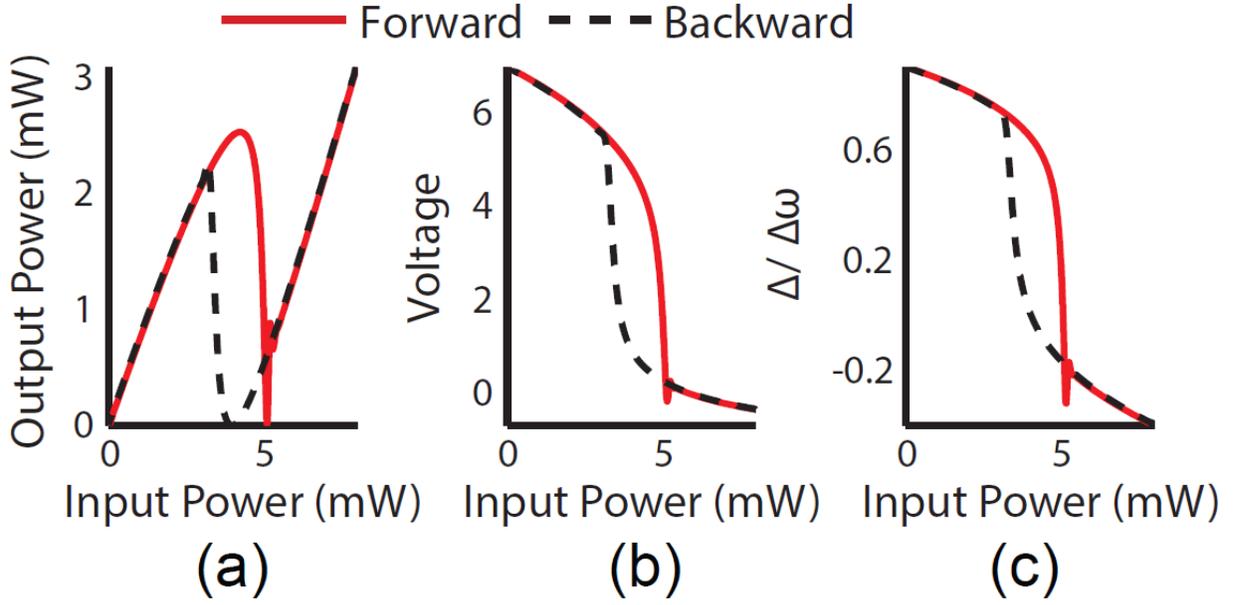

*Figure 3: (a) Cavity output power as a function of the input laser power for a self-electro-optic effect device, where the cavity modulation is achieved via the depletion capacitance of the diode, causing the cavity resonance to shift as a square root of the applied voltage. The parameters of the simulation are: $\frac{\Delta_o}{2\pi} = 200$ GHz; $\frac{\gamma_c}{2\pi} = 10$ GHz and $\frac{\gamma_l}{2\pi} = 10$ GHz; $V_o = 7$ V; $\alpha = 150$; $R = 1500$ Ω and $S = 1 A/W$. (b) The voltage across the diode and (c) the ring resonance shift, both as functions of the input power.*

**Symmetric self-electro-optic effect:** Although the optical bistability arises from the self-electro-optic effect, useful optical computing requires a slightly more complicated device known as a "symmetric" self-electro-optic bistable device. In this arrangement, we use two self-electro-optic bistable devices connected electrically in series (Figure 4a). Each of the devices acts as the load for the other, which can be controlled through exposure to light. In a symmetric self-electro-optic bistable device, the operating point of the device depends on the amount of light incident on the two diodes. The current-voltage relation for two diodes is given by:

$$I_{1,2} = I_o e^{-q(V_{1,2}+V_{bi})/kT} + I_{ph1,2}$$

where $I_{1,2}$ and $V_{1,2}$ are, respectively, the current and voltage across the two diodes, and $I_{ph1,2}$ is the photo-current in the diodes induced by the incident light on each of them ($P_{in1,2}$). The operating point is given by the intersection of the *I-V* curves for the two diodes (Figure 4b). When the power incident on each diode is different (a ratio of 1:2 for the optical power will suffice), then the two curves intersect only at a single point (shown by the blue circles in Figure 4b). This shows that depending on the ratio of input power each diode receives, most of the voltage will drop across one diode or the other. We note that when the ratio is equal (1:1), then there are three operating points (shown by the red circles). The center point of operation, where the voltage is equally distributed among the two diodes, is an unstable one, which will be crossed when the optical power incident on one diode changes from a low to high value. However, if the amount of optical power is very low, the generated photo-current is negligible, causing an equal voltage drop across both the diodes. This will ultimately decide the lowest power at which the switch can work. Figure 4c shows the voltage across each of the diodes as a function of the input optical power incident on diode 1 ($P_{in1}$), with the incident optical power on diode 2 kept constant at 1 mW. We also show the output powers from the two rings, as a function of $P_{in1}$ (Figure 4d). A clear bistable behavior is observed in the output power from both of the diodes.

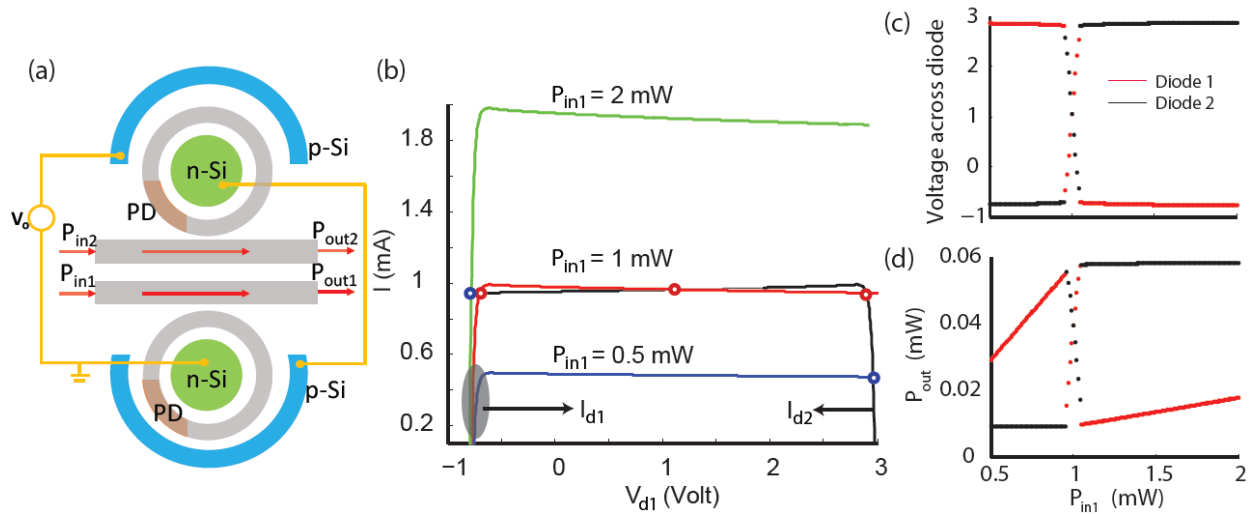

*Figure 4: (a) Schematic of a symmetric self-electro-optic bistable device, with two ring resonators connected electrically in series. (b) The current through diode 1 and diode 2 as a function of the voltage across diode 1, for different optical powers incident on diode 1 but with the power incident on diode 2 fixed at 1 mW. The intersections of the two curves give the operating points. We note that depending on the ratio of the input powers of the two diodes, we can have one or three points of operation. This gives rise to the bistability. (c) The voltage across the two diodes as a function of input power to diode 1, with input power to diode 2 fixed at 1 mW. (d) Output power from the two ring cavities as a function of the input power to diode 1, with input power to diode 2 fixed at 1 mW.*

**Platform for optical switch:** So far we have modeled the optical bistability in a silicon photonics platform with self-electro-optic (and subsequently symmetric self-electro-optic) devices. In this section, we analyze how a symmetric self-electro-optic device satisfies all the basic criteria[2] of an optical switch required to perform digital optical computing. In this device, we use a single polarization and frequency of light, and hence several devices can be cascaded easily, without any additional overhead. Also, due to the optical bistability, the signal is cleaned in every stage, so noise in the signal does not propagate through the system. Most optical switches based on an optical bistability satisfy these two criteria. However, these devices fail to demonstrate control-to-signal isolation; their logic levels are loss-dependent and they absolutely rely on critical biasing for achieving gain, a functionality required in order to achieve high fan-out.

The fundamentally different property of a symmetric self-electro-optic device, which enables the achievement of all the aforementioned features, is that this device is not only bistable in the optical power itself (as shown by the input-output plots in Figure 2 and Figure 3), but is also bistable in the ratio of the input and output optical powers. Hence one can encode a logical '0' ('1') by a low (high) ratio of the optical beams incident on or coming out of the two self-electro-optic devices. In fact, the device can be made bistable over a large range of optical powers (Figure 5a-c), just by changing the optical power incident on diode 2. If we plot the ratio of the output powers from the two diodes as a function of the ratio of their input powers (for each of the regimes in Figures 5a-c), we observe clear bistable operation in each regime. We emphasize that the bistability curves for three different power ranges look similar (Figure 5d), indicating that the device is capable of showing a new type of gain, known as time-sequential gain[3]. In a device exhibiting time-sequential gain, its transmission can be set by a low-power beam (the control beam). Then the control beams are turned off while signal beams are incident on the devices. Although these signal beams can be very high-power, their transmission is controlled by the state of the devices, set by low-power control beams. Thus, a low-power signal can control a high-power signal, and thus it exhibits a gain that is sufficient to realize large fan-out. We note that this gain does not require a critical biasing, as needed in other optically bistable devices. As explained above, the signal and control beams are present at different points of time, thus isolating the input from the output. Since the device is bistable with respect to the ratio of the optical powers, fluctuation of the input power does not degrade device performance as long as the light on both diodes is drawn from the same source. The logic level is again defined by the ratio between the two beams entering or exiting the two diodes, and hence both are equally affected by the propagation loss. Hence the logic level is independent of the loss. We note that the value for 'logic 1' is lower when the optical power is in the nanowatt range, leading to a lower contrast between the two logic levels. At a lower power this contrast diminishes further, ultimately making the device unusable. This happens because at a low optical power, the generated photo-current in the two diodes is small, leading to an equal voltage drop across the two diodes, as explained earlier.

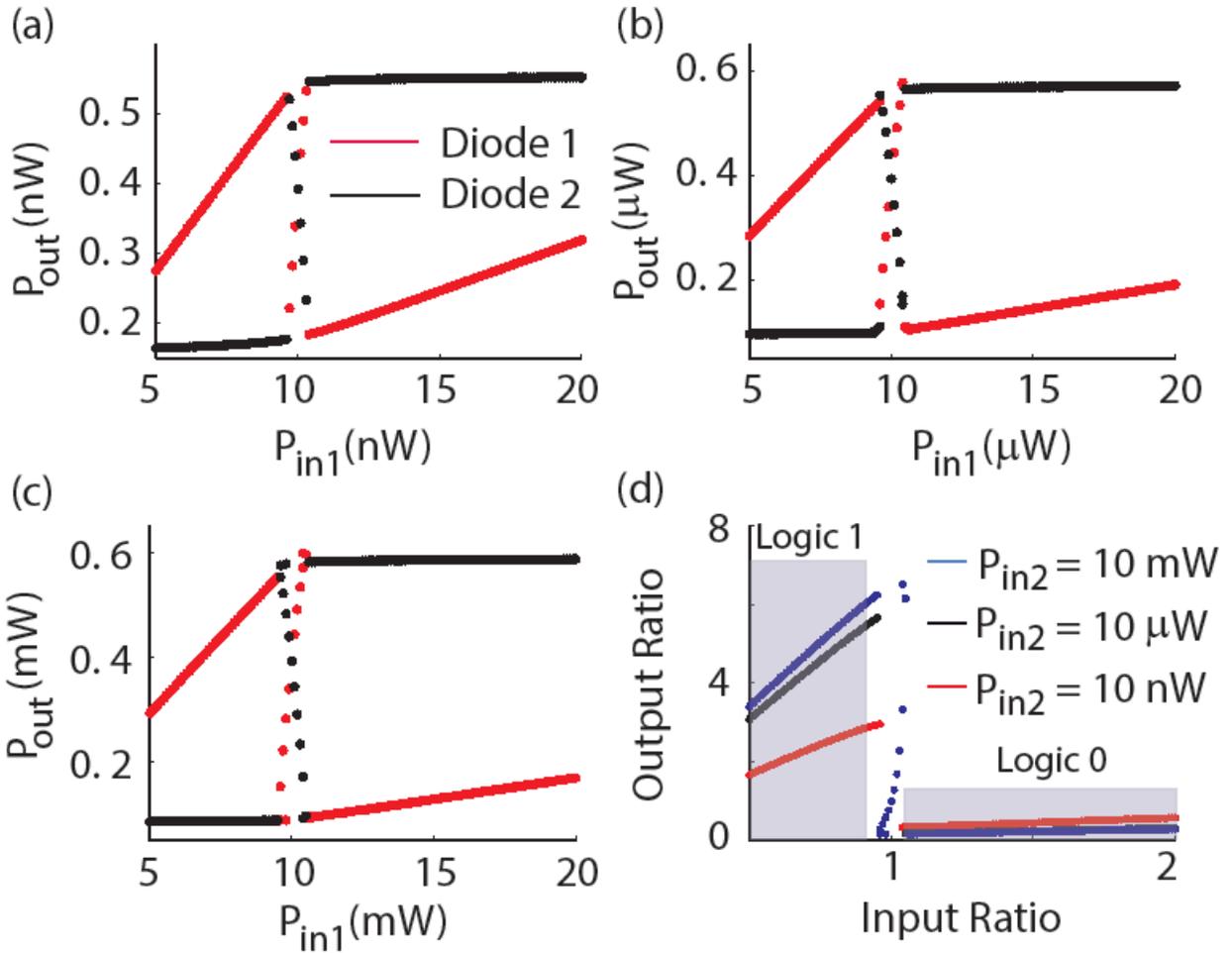

*Figure 5: Output power from the two ring cavities as a function of the input power to diode 1, with input power to diode 2 fixed at 10 nW (a), 10 µW (b), and 10 mW (c), for three different power ranges spanning six orders of magnitude. (d) The bistable operation solely depends on the power ratio, showing that it can be used to achieve time-sequential gain. The 'logic 1' and 'logic 0' regions are highlighted.*

So far we have analyzed the utility of self-electro-optic devices for conventional digital optical computing. At present, digital optical computing is not a competitive technology to electronic computing, and several other computing paradigms for optics are being explored, for example neuromorphic reservoir opto-electronic computing[23]. For such computing, one needs an S-shaped light in—light out curve, which can also be realized by means of a self-electro-optic bistability. Hence, these self-electro-optic devices will be beneficial in realizing not just digital optical computing, but also other regimes of optical computing.

**Energy Consumption & Speed Analysis:** Finally, we analyze the energy consumption and the speed of the self-electro-optic bistable device. The energy consumption of the device is given by the energy required to charge and discharge the depletion capacitance of the p-i-n diode. Note that successful operation of the self-electro-optic device requires efficient light absorption. In practice, an absorbing material is embedded in the ring, and the absorption strength of this region is specified in dB/μm. We find that at a wavelength of 1.5 μm, for a ring of radius 3 μm and an absorptive loss of ~0.05 dB/μm, the absorptive loss rate becomes $\frac{\gamma_l}{2\pi} = 12$ GHz, and the transmission $t = 0.9$ for critical coupling. Such performance can be achieved with state-of-the-art ring cavities and graphene-based waveguides[24, 25]. For a ring resonator with radius $r$, thickness $d$ and a depletion layer width of $w$, the capacitance $C_d$ is given by $C_d = 2\varepsilon_o \varepsilon_d \pi r d/w$, and so the switching energy is $E_{switch} = \frac{1}{2} C_d V_o^2$. With a ring radius of 3 μm, a ring slab thickness of 400nm, a depletion layer width of 1 μm, and an applied voltage of 4V (as derived from the simulations), we find the switching energy to be ~5 fJ. The electronic speed of operation will be limited by the speed of the feedback loop. For a resistance of ~2000 Ω, the electronics-limited speed of operation is ~100 GHz. However, the operational speed will also be limited by the photonic cavity quality factor to $(\gamma_c + \gamma_l)/\pi$ ~24 GHz.

**Conclusion:** We have analyzed the performance of a cavity-assisted self-electro-optic bistable device in a silicon photonics platform. The device can be made bistable at very low optical power, as the bistability is not caused explicitly by (comparatively weak) nonlinear effects. It is due to a positive opto-electronic feedback created by connecting an electro-optic modulator and a detector. The switching energy of the device can be quite low (~5fJ) while maintaining a speed of ~20GHz.

AUTHOR INFORMATION

A.M. conceived the idea and wrote the paper with help from A.R.

**Corresponding Author**

* To whom correspondence should be addressed. Email:arka@uw.edu

**Funding Sources**

ACKNOWLEDGMENT: A.M. acknowledges useful discussion with Feng Wang, Jelena Vuckovic and David Miller.

**Supplement**

**Relation between ring parameters and the general cavity model:**

For a ring modulator the transmission function is:

$$\frac{P_{out}}{P_{in}} = \frac{|t|^2 + \mu^2 + 2\mu|t|\cos(\theta_t - \theta_1)}{1 + |t|^2\mu^2 + 2\mu|t|\cos(\theta_t - \theta_1)}$$

where $\theta_1 = \frac{2\pi n L}{\lambda}$ signifies the phase accumulation in the ring, μ denotes the loss in the ring, and t is the transmission of the input light directly to the output. Note that when there is no round-trip loss (be it fixed loss or loss per unit length), μ equals unity[20]. At resonance, $\theta_t - \theta_1 = (2m + 1)\pi$. Let us denote this phase by θ. The critical condition is then given by $|t| = \mu$, and the transmission function is given by

$$\frac{P_{out}}{P_{in}} = \frac{|t|^2 + \mu^2 + 2\mu|t|\cos(\theta)}{1 + |t|^2\mu^2 + 2\mu|t|\cos(\theta)}$$

If we focus on the region close to a resonance, we can write

$$\theta = (2m+1)\pi - \theta' = \frac{2\pi n L}{\lambda} = \frac{2\pi n L}{\lambda_o + \Delta\lambda} = \frac{2\pi n L}{\lambda_o}\left(1 + \frac{\Delta\lambda}{\lambda_o}\right)^{-1} \approx \frac{2\pi n L}{\lambda_o} - \frac{2\pi n L \Delta\lambda}{\lambda_o^2}$$

Hence we can simplify the transmission function as:

$$\frac{P_{out}}{P_{in}} = \frac{|t|^2 + \mu^2 - 2\mu|t| - 2\mu|t|\cos\left(\frac{2\pi nL\Delta\lambda}{\lambda_o^2}\right)}{1 + |t|^2\mu^2 - 2\mu|t| - 2\mu|t|\cos\left(\frac{2\pi nL\Delta\lambda}{\lambda_o^2}\right)}$$

We can further expand the cosine term (as we are near the resonance) to write:

$$\frac{P_{out}}{P_{in}} = \frac{|t|^2 + \mu^2 - 2\mu|t| - 2\mu|t|\left(1 - \frac{1}{2}\left(\frac{2\pi nL\Delta\lambda}{\lambda_o^2}\right)^2\right)}{1 + |t|^2\mu^2 - 2\mu|t| - 2\mu|t|\left(1 - \frac{1}{2}\left(\frac{2\pi nL\Delta\lambda}{\lambda_o^2}\right)^2\right)}$$

This leads to

$$\frac{P_{out}}{P_{in}} = \frac{(|t| - \mu)^2 + \mu|t|\left(\frac{2\pi nL\Delta\lambda}{\lambda_o^2}\right)^2}{(1 - |t|\mu)^2 + \mu|t|\left(\frac{2\pi nL\Delta\lambda}{\lambda_o^2}\right)^2} = \frac{(|t| - \mu)^2 + \beta\Delta^2}{(1 - |t|\mu)^2 + \beta\Delta^2}$$

with

$$\beta = \frac{4\mu|t|\pi^2 n^2 L^2}{c^2}$$

Comparing terms we can write:

$$\gamma_c - \gamma_l = \frac{(|t| - \mu)}{\sqrt{\beta}}$$

and

$$\gamma_c + \gamma_l = \frac{(1 - |t|\mu)}{\sqrt{\beta}}$$

Combining these expressions we can write

$$\gamma_c = \frac{(1 + |t|)(1 - \mu)}{2\sqrt{\beta}}$$

$$\gamma_l = \frac{(1-|t|)(1+\mu)}{2\sqrt{\beta}}$$

Note that in experiments, the loss rates are measured in dB/m. For a loss rate of $p$ dB/m, we can write (for a ring length of L)

$$10 \log_{10} \mu = pL$$

Hence we can write the absorption linewidth as

$$\gamma_l = \frac{1}{4} \frac{(1-|t|)(1+\mu)}{\sqrt{\mu|t|}} \frac{c}{\pi n L}$$

We note that the linewidth explicitly depends on the ring length.